\documentclass[aps,prc,linenumbers,onecolumn,superscriptaddress, showkeys,nofootinbib]{revtex4}  

\usepackage[utf8]{inputenc}
\usepackage{graphicx,color}
\usepackage{epstopdf}
\usepackage{hyperref}

\usepackage{xspace}	

\newcommand{\GeVc} {\mbox{GeV/$\textit{c}$}\xspace}

\def  \vn        {\mbox{$\textit{v}_{n}$  }\xspace}

\def  \pT        {\mbox{$p_{\rm T}$  }\xspace}

\begin{document}

\title{Centrality and transverse momentum dependence of higher-order flow harmonics of identified hadrons in Au+Au collisions at $\sqrt{s_{\rm NN}}$ = 200~GeV}

\affiliation{Abilene Christian University, Abilene, Texas   79699}
\affiliation{AGH University of Science and Technology, FPACS, Cracow 30-059, Poland}
\affiliation{Alikhanov Institute for Theoretical and Experimental Physics NRC "Kurchatov Institute", Moscow 117218, Russia}
\affiliation{Argonne National Laboratory, Argonne, Illinois 60439}
\affiliation{American University of Cairo, New Cairo 11835, New Cairo, Egypt}
\affiliation{Brookhaven National Laboratory, Upton, New York 11973}
\affiliation{University of Calabria \& INFN-Cosenza, Italy}
\affiliation{University of California, Berkeley, California 94720}
\affiliation{University of California, Davis, California 95616}
\affiliation{University of California, Los Angeles, California 90095}
\affiliation{University of California, Riverside, California 92521}
\affiliation{Central China Normal University, Wuhan, Hubei 430079 }
\affiliation{University of Illinois at Chicago, Chicago, Illinois 60607}
\affiliation{Creighton University, Omaha, Nebraska 68178}
\affiliation{Czech Technical University in Prague, FNSPE, Prague 115 19, Czech Republic}
\affiliation{Technische Universit\"at Darmstadt, Darmstadt 64289, Germany}
\affiliation{ELTE E\"otv\"os Lor\'and University, Budapest, Hungary H-1117}
\affiliation{Frankfurt Institute for Advanced Studies FIAS, Frankfurt 60438, Germany}
\affiliation{Fudan University, Shanghai, 200433 }
\affiliation{University of Heidelberg, Heidelberg 69120, Germany }
\affiliation{University of Houston, Houston, Texas 77204}
\affiliation{Huzhou University, Huzhou, Zhejiang  313000}
\affiliation{Indian Institute of Science Education and Research (IISER), Berhampur 760010 , India}
\affiliation{Indian Institute of Science Education and Research (IISER) Tirupati, Tirupati 517507, India}
\affiliation{Indian Institute Technology, Patna, Bihar 801106, India}
\affiliation{Indiana University, Bloomington, Indiana 47408}
\affiliation{Institute of Modern Physics, Chinese Academy of Sciences, Lanzhou, Gansu 730000 }
\affiliation{University of Jammu, Jammu 180001, India}
\affiliation{Joint Institute for Nuclear Research, Dubna 141 980, Russia}
\affiliation{Kent State University, Kent, Ohio 44242}
\affiliation{University of Kentucky, Lexington, Kentucky 40506-0055}
\affiliation{Lawrence Berkeley National Laboratory, Berkeley, California 94720}
\affiliation{Lehigh University, Bethlehem, Pennsylvania 18015}
\affiliation{Max-Planck-Institut f\"ur Physik, Munich 80805, Germany}
\affiliation{Michigan State University, East Lansing, Michigan 48824}
\affiliation{National Research Nuclear University MEPhI, Moscow 115409, Russia}
\affiliation{National Institute of Science Education and Research, HBNI, Jatni 752050, India}
\affiliation{National Cheng Kung University, Tainan 70101 }
\affiliation{Nuclear Physics Institute of the CAS, Rez 250 68, Czech Republic}
\affiliation{Ohio State University, Columbus, Ohio 43210}
\affiliation{Institute of Nuclear Physics PAN, Cracow 31-342, Poland}
\affiliation{Panjab University, Chandigarh 160014, India}
\affiliation{Pennsylvania State University, University Park, Pennsylvania 16802}
\affiliation{NRC "Kurchatov Institute", Institute of High Energy Physics, Protvino 142281, Russia}
\affiliation{Purdue University, West Lafayette, Indiana 47907}
\affiliation{Rice University, Houston, Texas 77251}
\affiliation{Rutgers University, Piscataway, New Jersey 08854}
\affiliation{Universidade de S\~ao Paulo, S\~ao Paulo, Brazil 05314-970}
\affiliation{University of Science and Technology of China, Hefei, Anhui 230026}
\affiliation{Shandong University, Qingdao, Shandong 266237}
\affiliation{Shanghai Institute of Applied Physics, Chinese Academy of Sciences, Shanghai 201800}
\affiliation{Southern Connecticut State University, New Haven, Connecticut 06515}
\affiliation{State University of New York, Stony Brook, New York 11794}
\affiliation{Instituto de Alta Investigaci\'on, Universidad de Tarapac\'a, Arica 1000000, Chile}
\affiliation{Temple University, Philadelphia, Pennsylvania 19122}
\affiliation{Texas A\&M University, College Station, Texas 77843}
\affiliation{University of Texas, Austin, Texas 78712}
\affiliation{Tsinghua University, Beijing 100084}
\affiliation{University of Tsukuba, Tsukuba, Ibaraki 305-8571, Japan}
\affiliation{United States Naval Academy, Annapolis, Maryland 21402}
\affiliation{Valparaiso University, Valparaiso, Indiana 46383}
\affiliation{Variable Energy Cyclotron Centre, Kolkata 700064, India}
\affiliation{Warsaw University of Technology, Warsaw 00-661, Poland}
\affiliation{Wayne State University, Detroit, Michigan 48201}
\affiliation{Yale University, New Haven, Connecticut 06520}

\author{M.~S.~Abdallah}\affiliation{American University of Cairo, New Cairo 11835, New Cairo, Egypt}
\author{B.~E.~Aboona}\affiliation{Texas A\&M University, College Station, Texas 77843}
\author{J.~Adam}\affiliation{Brookhaven National Laboratory, Upton, New York 11973}
\author{L.~Adamczyk}\affiliation{AGH University of Science and Technology, FPACS, Cracow 30-059, Poland}
\author{J.~R.~Adams}\affiliation{Ohio State University, Columbus, Ohio 43210}
\author{J.~K.~Adkins}\affiliation{University of Kentucky, Lexington, Kentucky 40506-0055}
\author{G.~Agakishiev}\affiliation{Joint Institute for Nuclear Research, Dubna 141 980, Russia}
\author{I.~Aggarwal}\affiliation{Panjab University, Chandigarh 160014, India}
\author{M.~M.~Aggarwal}\affiliation{Panjab University, Chandigarh 160014, India}
\author{Z.~Ahammed}\affiliation{Variable Energy Cyclotron Centre, Kolkata 700064, India}
% ...

\collaboration{STAR Collaboration}\noaffiliation


\begin{abstract}
We present high-precision measurements of elliptic, triangular,  and quadrangular flow $v_{2}$, $v_{3}$, and $v_{4}$,  respectively, at midrapidity ($|\eta|<1.0$) for identified hadrons $\pi$, $p$, $K$, $\varphi$, $K_s$, $\Lambda$ as a function of centrality and transverse momentum in Au+Au collisions at the center-of-mass energy $\sqrt{s_{\rm NN}}=$ 200 GeV. We observe similar $v_{n}$ trends between light and strange mesons which indicates that the heavier strange quarks flow as strongly as the lighter up and down quarks. The number-of-constituent-quark scaling for $v_{2}$, $v_{3}$, and $v_{4}$ is found to hold within statistical uncertainty for 0-10$\%$,  10-40$\%$ and 40-80$\%$ collision centrality intervals. The results are compared to several viscous hydrodynamic calculations with varying initial conditions, and could serve as an additional constraint to the development of  hydrodynamic models.
\end{abstract}

\maketitle


\section{Introduction}

A main goal of {\color{black}high-energy heavy-ion} facilities such as the Relativistic Heavy-Ion Collider (RHIC) and the Large Hadron Collider (LHC) is to understand the properties of the quark-gluon plasma (QGP)~\cite{Shuryak:1978ij,Shuryak:1980tp,Muller:2012zq}. Of particular importance are the transport properties of the QGP, especially the specific shear viscosity per unit of entropy density,  ($\eta/s$), which describes the ability of the QGP to transport and dissipate momentum. 
Anisotropic flow measurements quantify the azimuthal anisotropy of the particle emission in the transverse plane. These reflect the viscous hydrodynamic response to the initial spatial distribution of energy density produced in the early stages of the collision~\cite{Danielewicz:1998vz,Ackermann:2000tr,Gardim:2012yp,Lacey:2013eia,Adcox:2002ms,Heinz:2001xi,Hirano:2005xf,Huovinen:2001cy,Hirano:2002ds,Romatschke:2007mq,Luzum:2011mm,Song:2010mg,Qian:2016fpi,Schenke:2011tv,Teaney:2012ke}.

{\color{black}
Experimentally, anisotropic flow can be characterized using the Fourier expansion~\cite{Voloshin:1994mz,Poskanzer:1998yz} of the azimuthal distribution as~\cite{Poskanzer:1998yz}:
\begin{eqnarray}
\label{eq:1-1}
E \frac{d^{3}N}{d^3p} &=& \frac{1}{2\pi} \frac{d^2N}{p_{T} dp_{T} dy} \left( 1 +  \sum^{N}_{i=1} 2 \textit{v}_{n}  \cos\left(    n \left( \phi - \psi_{RP} \right)    \right)        \right),
\end{eqnarray}
where  \vn is the $n^{th}$ order flow coefficient, $E$ is the energy, $p_{T}$ is transverse momentum, $y$ is rapidity,  $\phi$ is the particle azimuthal angle, and $\psi_{RP}$ is the azimuth of the reaction plane given by the beam direction and impact parameter.
The first, second, third and fourth Fourier harmonics ($v_{1}$, $v_{2}$, $v_{3}$ and $v_{4}$) are called the directed flow, elliptic flow~\cite{Voloshin:1994mz,Poskanzer:1998yz}, triangular flow and quadrangular flow, respectively.
}

%
%

Previous measurements of identified hadrons by the STAR collaboration~\cite{Adams:2005zg,Adamczyk:2016gfs,Adamczyk:2015ukd,STAR:2004jwm,STAR:2003wqp} were limited to the elliptic flow $v_{2}$ and little information was shown about the higher-order flow harmonics $v_{n}$ with n$>$2. Those $v_{2}$($p_T$) {\color{black}($p_T = \sqrt{p_x^2 + p_y^2}$)} measurements showed mass-order dependence at low $p_T$, $p_T$ $<$ 2.0~GeV/$c$, which is understood to result from the hydrodynamic expansion of the medium~\cite{Singha:2016aim}. For the intermediate-$p_T$ region, 2.0 $<$ $p_{T}$ $<$ 4.0~GeV/$c$, the identified hadron $v_{2}$($p_T$) magnitudes are larger for baryons than mesons (which is referred to as baryon-meson splitting). Such an observation can be described by quark coalescence models~\cite{Voloshin:2002wa,Molnar:2003ff,Fries:2008hs}. In the quark coalescence picture, partons develop flow during the partonic evolution and the hadron flow is given by the sum of the collective flows of the constituent partons. The quark coalescence mechanism explains the observed number-of-constituent-quark (NCQ) scaling of $v_{2}$($p_T$) at RHIC.

In this paper, we extend the prior measurements by adding results on $v_3$ and $v_4$ of identified hadrons {\color{black}$\pi$, $p$, $K$, $\varphi$, $K_s$, $\Lambda$} for Au+Au collisions at $\sqrt{s_{\rm NN}}=200$~GeV as a function of both transverse momentum ($p_T$) and centrality. Due to the strong viscous effects on the higher-order anisotropic flow coefficients $v_{n}$ with $n > 2$, higher order harmonics $v_{n>2}$ are expected to be more sensitive to $\mathrm{\eta/s}$ than the elliptic flow $v_{2}$~\cite{Alver:2010dn,Teaney:2012ke}.
In addition, previous studies indicate that NCQ scaling ($n_q$ is the number of constituent quarks) works well for the elliptic flow $v_2$, but does not for the higher harmonics \cite{Adamczyk:2013gw,ALICE:2016cti}. {\color{black}As proposed in} Ref. ~\cite{Han:2011iy}, a modified form of the scaling function, $v_{n}/n_{q}^{n/2}$, is tested here. It works better for $v_3$ and $v_4$ up to the intermediate-$p_{T}$  region~\cite{Abelev:2007rw,Abelev:2010tr,Adams:2005zg,Abelev:2007qg,Voloshin:2008dg,STAR:2017ykf,PHENIX:2014uik}. Although hadronic rescattering might be treated as a reason of the modification in scaling, the underlying physics {\color{black}is under discussion}~\cite{Zhang:2015skc, Lacey:2011ug}.
%


The present measurements will not only supplement other anisotropic flow studies of identified particles for Pb+Pb collisions
at the LHC energies as reported in Refs~\cite{ALICE:2018yph,ALICE:2016cti,ALICE:2019xkq}, but also be compared to two hydrodynamic models~\cite{Alba:2017hhe,Schenke:2019ruo}, which are summarized in Table~\ref{tab:2}. The first, Hydro-1~\cite{Alba:2017hhe}, employs the TRENTO model~\cite{Moreland:2014oya} initial-state and does not include a hadronic afterburner. The second, Hydro-2~\cite{Schenke:2019ruo}, uses an IP-Glasma~\cite{Schenke:2012wb} initial-state in conjunction with a UrQMD~\cite{Bass:1998ca,Bleicher:1999xi} afterburner. Hydro-II also imposes the effects of global momentum and local charge conservation.

\begin{table*}[th]
\begin{center}
 \begin{tabular}{|c|c|c|}
 \hline
 $   $                      &     Hydro$-1$~\cite{Alba:2017hhe}         &  Hydro$-2$~\cite{Schenke:2019ruo} \\
 \hline
 $\eta/s$                   &          0.05                             &               0.12                                   \\
 \hline
 Initial conditions         &         TRENTO                            &          IP-Glasma                                 \\
 \hline
  Contributions             &         Hydro +                           &          Hydro +                                    \\
                            &         Direct decays                     &         Hadronic cascade                       \\
 \hline
\end{tabular} 
\caption{Summary of the two hydrodynamic models Hydro$-1$~\cite{Alba:2017hhe} and Hydro$-2$~\cite{Schenke:2019ruo}.
\label{tab:2}}
\end{center}
\end{table*}

This paper is organized as follows. Section~\ref{Sec:2} describes the experimental setup. In Section~\ref{Sec:3}, the particle identification, the event plane reconstruction, $v_{n}$ signal extraction, and systematic uncertainty estimation are discussed. In Section~\ref{Sec:4}, the centrality and momentum dependent $v_{n}$ results are presented and discussed. The summary is presented in Section~\ref{Sec:5}.

\section{Experimental setup} 
\label{Sec:2}
The Solenoidal Tracker At RHIC (STAR) at the Brookhaven National Laboratory employs a solenoidal magnet and multiple detectors to provide a wide-acceptance measurement at mid-rapidity \cite{Ackermann:2002ad}. In this analysis, the primary detectors used were the STAR Time-Projection Chamber (TPC) and the Time-of-Flight (ToF) systems. 

The TPC has a pseudorapidity, $\eta$, acceptance of $|\eta|$ $<$ 1.8, and full azimuthal coverage~\cite{Anderson:2003ur}. Along the beam direction, the central membrane divides the TPC into two halves. Within the TPC radius of $0.5 < r < 2$~m, tracks can be reconstructed with a maximum of 45 hit points per track. The specific energy loss ($dE/dx$) provided by the TPC for each reconstructed track can be used for particle identification. The time-of-flight detector is based on Multi-gap Resistive Plate Chambers (MRPCs)~\cite{Llope:2005yw}. The ToF has a time resolution of $\sim$85 ps, and covers the full azimuth and a pseudorapidity range of $|\eta|$ $<$  0.94. The particle mass-squared, ${\it m}^{2}$, provided by the ToF system significantly extends STAR's particle identification capabilities to higher $p_T$. Additional details on the use of these detectors
are provided in Section~\ref{Sec:3-a}.


The Au+Au 200 GeV data collected in the year 2011 {\color{black} with about 400~M events} is used in this analysis. 
A minimum bias trigger based on a coincidence of the signals from the Zero Degree Calorimeters (ZDC)\cite{Xu:2016alq}, Vertex Position Detectors (VPD)\cite{LLOPE201423}, and/or Beam-Beam Counters (BBC)\cite{Ackermann:2002ad} was used.
Collisions more than $\pm 30$~cm from the center of STAR along the beam direction, or more than 2 cm radially from the center of the beam pipe, were rejected.
The absolute difference between the $\textit{z}$-vertex positions measured by the TPC and VPD detectors in each event was required to be less than $3$ cm to reduce background events. Collision centrality is inferred from the measured event-by-event multiplicity with the aid of a Monte Carlo Glauber simulation~\cite{Alver:2008aq,Adamczyk:2012ku}.
%
Also, a multivariate quality assurance of each data-taking run was performed. The values of the mean transverse momentum, the mean vertex position,  and the mean multiplicity in the detector in single data-taking runs were all required to be within 3$\sigma$ away of their mean values over the entire data set. 
Track quality cuts were applied to suppress backgrounds and to improve the resolution of track quantities such as the momentum and energy loss. Each track was required to have at least 15 hits assigned to it (out of up to 45). In order to remove track splitting the ratio of the number of reconstructed hits to the maximum possible number of hits for each track was required to be larger than 0.51. Tracks with $|\eta|$ $>$ 0.9 and momenta below 0.2 GeV/$c$, or above 4.0 GeV/$c$, were rejected.

\section{Methodology} 
\label{Sec:3}

\subsection{Particle identification}
\label{Sec:3-a}

Particle identification in the STAR experiment can be done in multiple ways~\cite{Adamczyk:2013gw}.
The identification of charged particles is based on a combination of momentum information, the specific energy loss $dE/dx$ in the TPC and a required time-of-flight measurement with the ToF detector. 
Charged pions and kaons can be easily distinguished on the basis of their $dE/dx$ values for momenta up to approximately $0.7$ GeV/$c$; at higher momenta the particles' $dE/dx$ distributions overlap. At higher momenta, two-dimensional fits in a combined $m^{2}$ vs. $dE/dx$ plane were used to statistically extract the particle yield for $\pi$ and $K$~\cite{Adamczyk:2013gw}. 
Protons and antiprotons are identified mainly by the time-of-flight $m^{2}$ information. To suppress contributions from pions and kaons an additional cut on $|n\sigma_{p}| < 2.5$ was applied. 
At low transverse momenta ($p_{T} < $ 2 GeV/$c$) the separation of protons relative to pions and kaons is good enough to count all protons within an equivalent range of 3$\sigma$ around the center of the $n\sigma_{p}$ distribution.
At high $p_{T}$ the tails on the left of the proton distributions are excluded to avoid contamination from pions and kaons. Thus the $m^{2}$ cut value increases with $p_{T}$.

The unstable particles $K^{0}_{s}$, $\varphi$, $\Lambda$, and $\bar{\Lambda}$  decay into a pair of oppositely charged particles and can be reconstructed using the invariant mass technique.
For weak decay particles, additional topological constraints on the decay kinematics were applied to suppress backgrounds. 
The combinatorial background from uncorrelated particles was reduced by employing cuts on the daughter particle $dE/dx$ and/or $m^{2}$, as well as on the topology of the specific decay.
The misidentification of the daughter particles, which is more probable at higher momenta, can result in an additional correlated background. 
Such a correlated background, for example from the $\Lambda$ hyperon, can appear in the $\pi^{+}\pi^{-}$($K^{0}_{S}$) invariant mass distribution if the proton was misidentified as a $\pi^{+}$. 
Such a correlated background does not create a peak in the invariant mass distribution of the particles of interest because the daughter-particle masses are assumed to be the nominal ones (e.g., $\pi$ mass instead of proton mass), but instead appears as a broad distribution which can significantly affect the signal extraction. 
This correlated background can be eliminated by investigating additional invariant mass spectra with identical track combinations, but different daughter mass values.
The background was removed by applying invariant mass cuts on the corresponding unwanted peaks in the misidentified invariant mass distributions. Usually, the correlated background from particle misidentification increases with the $p_{T}$ values of the mother particle. In this work, the remaining uncorrelated combinatorial background was subtracted with the mixed-event technique.

\subsection{$v_{n}$ Analysis method}  

In this work we used the two-particle cumulant method to extract the flow coefficients $v_{n}$ of $\pi$, $K$, and $p$. For other particles we used the event plane (EP) method to measure the $v_{2}$ and $v_{3}$ values. In this section, a description of each method used here is provided.

\subsubsection{Two-particle cumulant method}
\label{Sec:3-1}

The framework for the cumulant method is described in Refs.~\cite{Bilandzic:2010jr,Bilandzic:2013kga}, which was extended to the case of subevents in Refs.~\cite{Gajdosova:2017fsc,Jia:2017hbm}. The two-particle correlations were constructed using the two-subevent cumulant method~\cite{Jia:2017hbm}, with particle weights, e.g. weighted with the particle acceptance correction, and $\Delta\eta~ > 0.7$ separation between the subevents $\textit{A}$ and $\textit{B}$ (i.e. $1> \eta_{A} > 0.35$ and $-1< \eta_{B} < -0.35$). The use of the two-subevent method reduces the nonflow correlations including the decay of resonances to several charged daughter particles, Hanbury-Brown Twiss correlations, and jets~\cite{Voloshin:2008dg}. The two-particle {\color{black}flow harmonics} can be written as,
%
\begin{eqnarray}\label{eq:2-1}
v^{2}_{n} &=&   \langle  \langle \cos (n [\varphi^{A}_{i} -  \varphi^{B}_{j} ] ) \rangle \rangle,
\end{eqnarray}
where $\langle \langle \, \rangle \rangle$ indicates the average over all particles in a single event and over all events, and $\varphi_{i}$ is the azimuthal angle of the i$^{th}$ particle. The integrated and $p_{\mathrm{T}}$ differential n$^{th}$-order flow harmonics are given as,
\begin{eqnarray}\label{eq:2-1x}
v_{n}      &=&  \langle  \langle \cos (n [\varphi^{A}_{i} -  \varphi^{B}_{j} ] ) \rangle \rangle/\sqrt{v^{2}_{n}},
\end{eqnarray}
and 
\begin{eqnarray}\label{eq:2-1x}
v_{n}(\pT, PID) &=&  \langle  \langle \cos (n [\varphi^{A}_{i}(\pT, PID) -  \varphi^{B}_{j} ] ) \rangle \rangle/\sqrt{v^{2}_{n}}.
\end{eqnarray}

\subsubsection{Event plane method}
\label{Sec:4-2}
The n$^{th}$-order event planes, $\Psi_{n}$, used here are constructed from the azimuthal distribution of final-state particles~\cite{Poskanzer:1998yz} as
\begin{equation}
\Psi_{n} = \tan^{-1} \left( \frac{\sum_{i}w_{i}\sin(n\varphi_{i})}{\sum_{i}w_{i}\cos(n\varphi_{i})} \right)/n
\label{form_PsiAngle}
\end{equation}
where $\varphi_{i}$ is the azimuthal angle of i$^{th}$ particle and $w_{i}$ is its weight that reflect the detector $\eta$-$\phi$ acceptance correction. Only tracks with momentum in the range from 0.2 to 2 \GeVc and pseudorapidity $|\eta| < 1$ in the TPC were used to calculate the event plane(s).

Two planes (east and west) are constructed using tracks from the opposite pseudorapidity hemisphere to the particle of interest, i.e., the east $\eta_{sub}$ event plane using tracks with $-1.0 \leq \eta \leq -0.05$ and the west $\eta_{sub}$ event plane using tracks with $0.05 \leq \eta \leq 1.0$. This procedure is called ''$\eta$-sub'' method and suppresses the nonflow contribution ~\cite{Voloshin:2008dg}. The additional bias in the event plane reconstruction caused by detector inefficiencies generates a non-uniform $\Psi_{n}$ angle distribution in the laboratory coordinate system.
To flatten this distribution, 
the recentering ~\cite{Selyuzhenkov:2007zi} and shifting ~\cite{Barrette:1997pt} method were applied. The event plane resolution was calculated from the two $\eta$-sub events~\cite{Voloshin:2008dg}. 
Each of the flow harmonics was measured with respect to the corresponding, same-order, event plane.

\subsection{Systematic uncertainty analysis} 
\label{Sec:3Systematics}

The systematic uncertainties associated with the measurements shown in this paper are evaluated by varying several parameters of the analysis and comparing the measurements with their nominal values.
The systematic uncertainty correlated with the event selection is evaluated by studying the variation of the results with different selections on the primary vertex position, i.e. using a range $-30$ to $0$~cm or $0$ to $30$~cm rather than the nominal range of $\pm 30$~cm.  The event-cuts systematic uncertainty ranges from 1\% to 2\% from central to peripheral collisions.
 The systematic uncertainty resulting from the track selection is estimated by applying stricter conditions: (i)   DCA is reduced to be less than 2~cm rather than 3~cm, and (ii) the number of TPC space points changing from more than $15$ points to more than $20$ points.
 The track-cuts systematic uncertainty ranges from 1\% to 3\% from central to peripheral collisions.
The systematic uncertainty correlated with the nonflow correlations due to
 Bose-Einstein correlations,  resonance decays, and the fragments of individual jets is evaluated by varying the pseudorapidity gap, $\Delta\eta~=~\eta_{1}-\eta_{2}$, for the track pairs used in the measurement. The variation of the results for $\Delta\eta$ values of 0.6 and 0.8 was studied.
The systematic uncertainty from the nonflow correlations ranges from 2\% to 5\% from central to peripheral collisions.
%
{\color{black}The systematic uncertainty from varying the particle identification cuts about their nominal values~\cite{Adamczyk:2013gw} ranges from 1\% to 3\% from central to peripheral collisions.}
The overall systematic uncertainty, considering all sources as independent of each other, was evaluated via the quadrature sum of the uncertainties from the individual cut variations. They range from 3\% to 7\% from central to peripheral collisions.


\section{Results and discussion}
\label{Sec:4}

The $v_n$ results for Au+Au collisions at 200 GeV are shown as a function of the transverse momentum in Sec. \ref{Sec:4-1}, kinetic energy in Sec. \ref{Sec:4-2} and centrality in Sec. \ref{Sec:4-3}. 
The statistical uncertainties are shown as the straight vertical lines, while the point-by-point systematic uncertainties are shown as the open boxes.  

\subsection{$v_{n}$ as a function of transverse momentum}
\label{Sec:4-1} 
\begin{figure*}[th]
\vskip -0.36cm
\centering{
\includegraphics[width=1.02 \linewidth,angle=0]{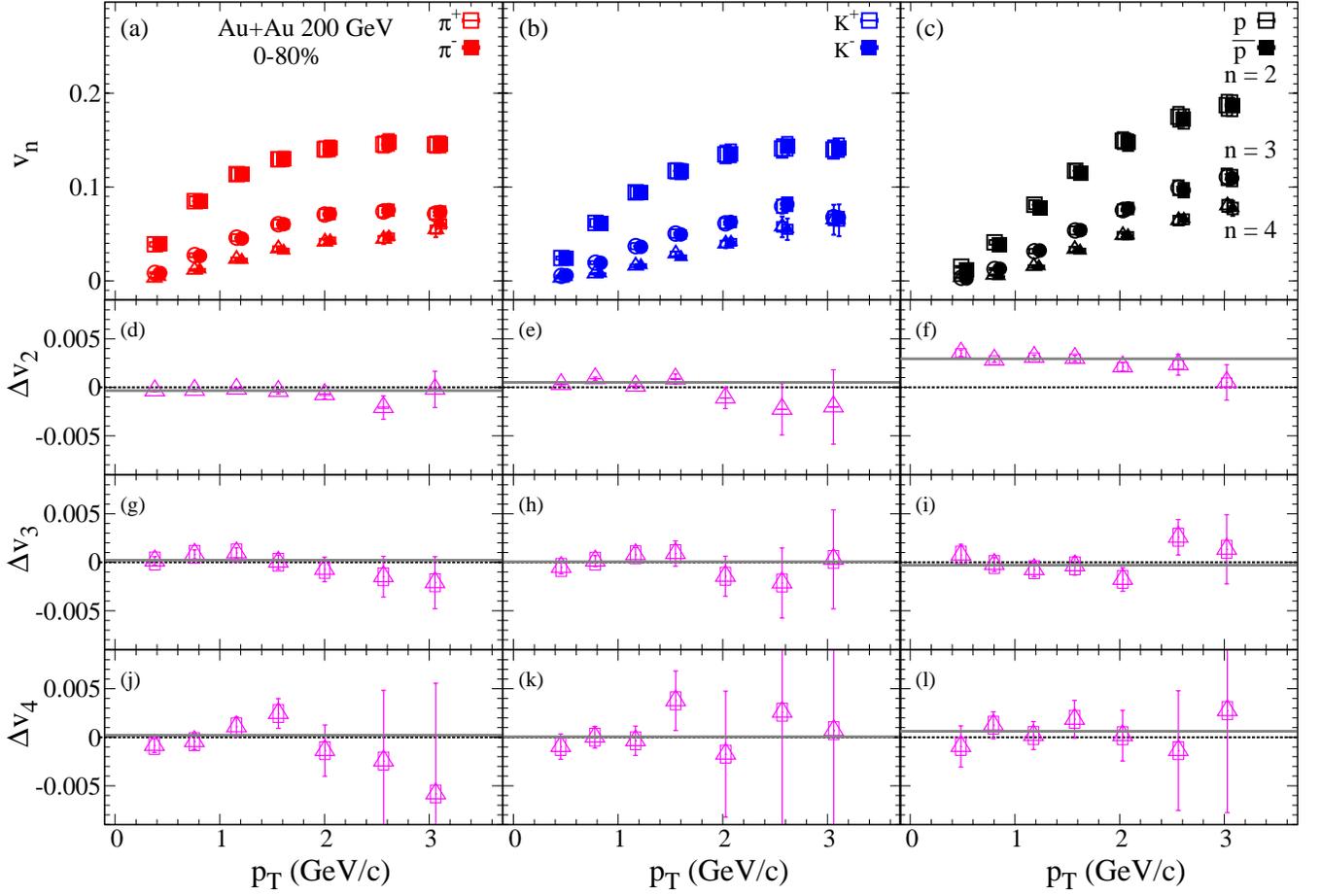}
\vskip -0.36cm
\caption{
The panels (a)-(c) show the transverse momentum dependence of elliptic, triangular and quadrangular flow of particle and antiparticle for 0--80\% central Au+Au collisions at $\sqrt{s_{\rm NN}}$ = 200~GeV  using the two-particle cumulant method. The panels (d)-(l) represent the $v_{2}$, $v_{3}$, and $v_{4}$ difference between positive and negative particles. Solid lines are linear fits to the data.
\label{fig:fig1}
 }
}
\vskip -0.2cm
\end{figure*}

The panels (a)-(c) of Fig.~\ref{fig:fig1} present the 
 particle and antiparticle $v_2$, $v_3$ and $v_4$ for 0--80\% central Au+Au collisions at $\sqrt{s_{\rm NN}}$ = 200~GeV. The measurements show clear similarities in the values and trends between the particle and antiparticle. A more quantitative conclusion can be made by forming the differences $\Delta v_2$,  $\Delta v_3$  and $\Delta v_4$ which are shown in panels (d)-(i). The $\Delta v_2$~\cite{Adamczyk:2013gw}, $\Delta v_3$ and $\Delta v_4$ values for pions and kaons indicate little if any difference between positive and negative mesons of the same species. 
Although the $\Delta v_3$ and $\Delta v_4$ show little if any difference between protons and antiprotons,  the $\Delta v_2$ is nonzero, with a value of 0.0028 $\pm$ 0.0002 (stat) $\pm$ 0.0003(syst). As pointed out in our prior studies~\cite{STAR:2013cow,STAR:2015rxv} the $v_n$ difference between positive and negative particles could be accounted for by considering nuclear stopping power which decreases with increasing $\sqrt{s_{\rm NN}}$. Such an effect is expected to be small at $\sqrt{s_{\rm NN}}$ = 200~GeV.

\begin{figure}[th]
\vskip -0.36cm
\centering{
\includegraphics[width=1.01 \linewidth,angle=0]{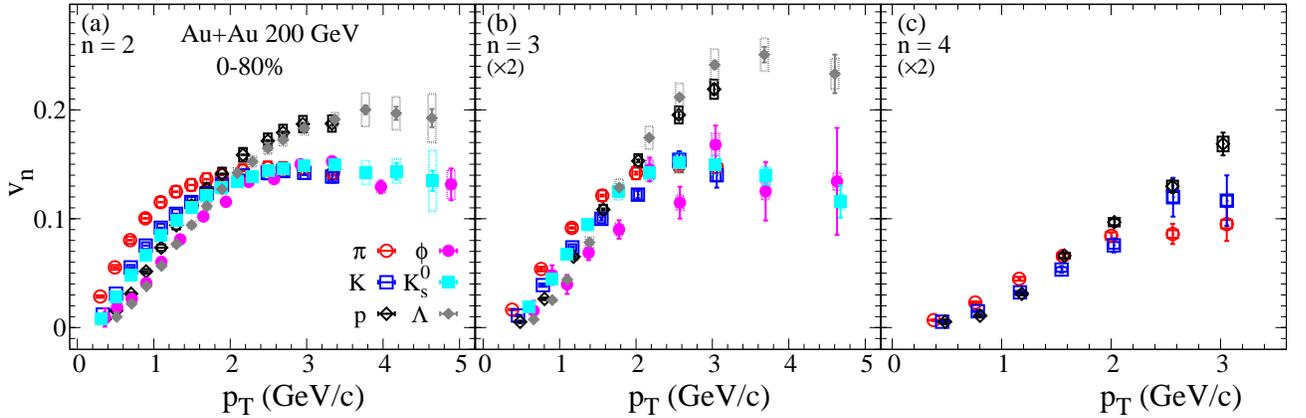}
\vskip -0.36cm
\caption{
The transverse momentum dependence of the identified particle $v_2$ (a), $v_3$ (b) and $v_4$ (c) for 0--80\% central Au+Au collisions at $\sqrt{s_{\rm NN}}$ = 200 GeV.
\label{fig:fig2}
 }
}
\vskip -0.2cm
\end{figure}

Figure~\ref{fig:fig2} shows the transverse momentum dependence  at midrapidity of $v_2$ (a) and $v_3$ (b) of $\pi$, $K$, $p$,  $\Lambda$, $\varphi$ and $K_{s}^{0}$ and 
$v_4$ (c) of $\pi$, $K$ and $p$
for 0--80\% central Au+Au collisions at $\sqrt{s_{\rm NN}}$ = 200~GeV. The measurements indicate similar increasing then flattening trends {\color{black}as a function of $p_T$} in $v_{n=2,3,4}(p_{T})$ for all particles shown. Also mass ordering at low $p_T$ is observed for $v_2$, $v_3$, and $v_4$. The shapes of the flow harmonics for light and strange mesons are comparable, which suggests similar flow strength for u, d, and s quarks. 
\begin{figure*}[th]
\centering{
\includegraphics[width=0.95 \linewidth,angle=0]{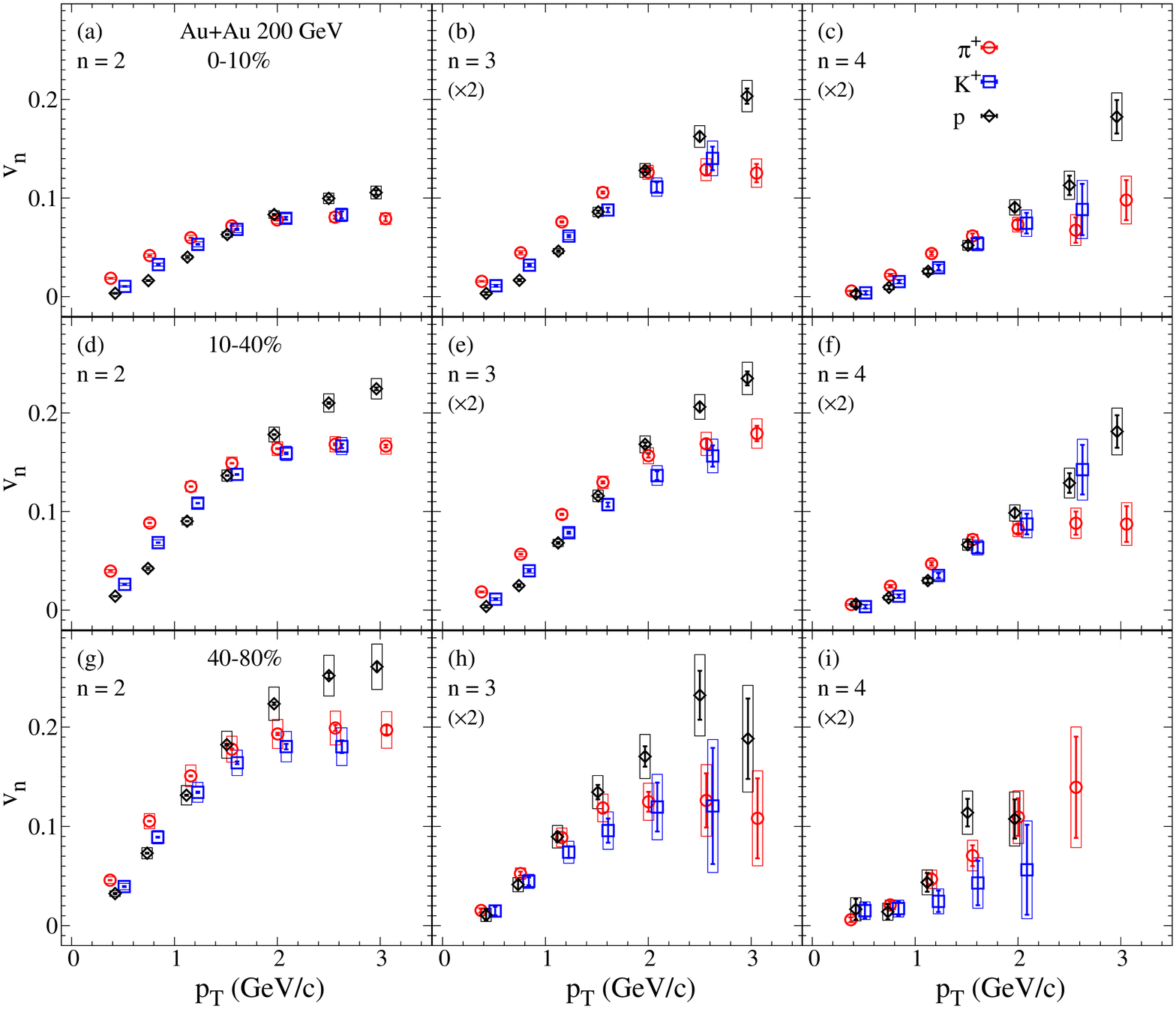}
\caption{The transverse momentum dependence of the identified particle $v_{2}$, $v_{3}$, and $v_{4}$ for 0--10\%, 10--40\% and 40--80\% central Au+Au collisions at $\sqrt{s_{\rm NN}}$ = 200 GeV.
\label{fig:fig3p}}
}
\end{figure*}
\begin{figure*}[th]
\centering{
\includegraphics[width=0.9 \linewidth,angle=0]{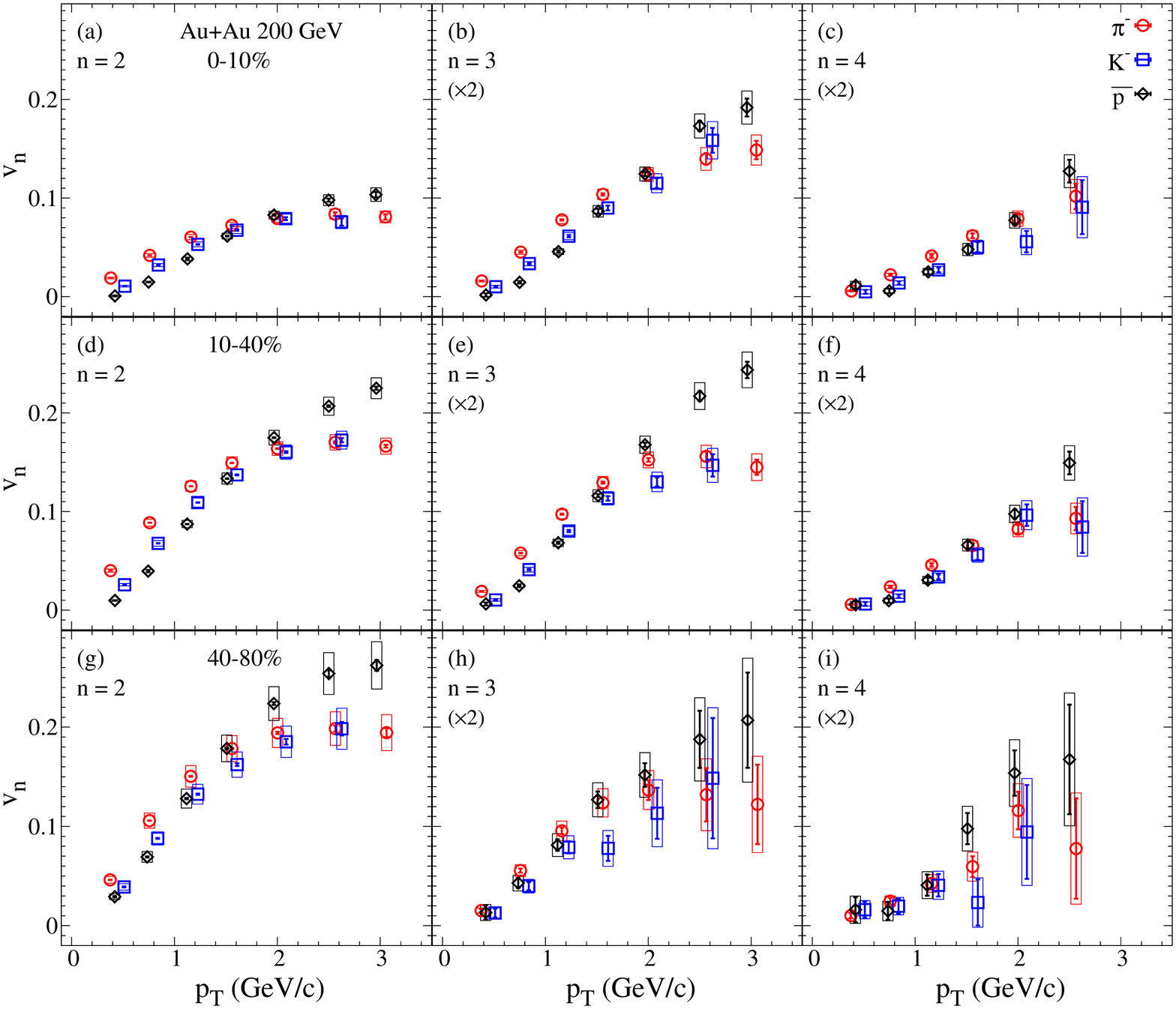}
\caption{The transverse momentum dependence of the identified antiparticle $v_{2}$, $v_{3}$, and $v_{4}$ for 0--10\%, 10--40\% and 40--80\% central Au+Au collisions at $\sqrt{s_{\rm NN}}$ = 200 GeV.
\label{fig:fig3m}}
}
\end{figure*}

The $p_T$ dependence of  $v_2$, $v_3$, and $v_4$  for $\pi$, $K$, $p$ and their charge conjugates are shown in Figs.~\ref{fig:fig3p} and~\ref{fig:fig3m}. The measurements indicate mass ordering at low $p_T$ for $v_2$, $v_3$, and $v_4$. Our measurements are in good agreement with the prior measurements~\cite{Adamczyk:2015ukd,STAR:2008ftz,PHENIX:2014uik}.
The $v_{2}$ values are found to be higher in peripheral collisions (40-80$\%$ centrality) compared to those in central collisions (0-10$\%$ centrality). The $v_{3}$ and $v_{4}$ values indicate a weak centrality dependence. This observation is compatible with the picture in which the viscous effects reduce the initial spatial anisotropic effects on the higher-order flow harmonics~\cite{Lacey:2013eia,Schenke:2014tga}.

\subsection{Scaled $v_{n}$ as a function of scaled kinetic energy}
\label{Sec:4-2}


Prior investigations~\cite{STAR:2017ykf,STAR:2015rxv} have indicated that particle species dependence remains in plots of $v_n$ versus $p_T$ when each are scaled by the number of constituent quarks, $n_q$.  The breakdown of this scaling is also shown for the present data in Appendix~\ref{A1}. 
A modified scaling function of $v_{n}/n_{q}^{n/2}$ vs  scaled kinetic energy ($KE_{T} = m_{T} - m_{0}$ and $m_{T} = \sqrt{p_{T}^{2}+m_{0}^{2}}$)  is suggested to work better, and will be tested in this work.

\begin{figure}[th]
\vskip -0.36cm
\centering{
\includegraphics[width=0.70 \linewidth,angle=0]{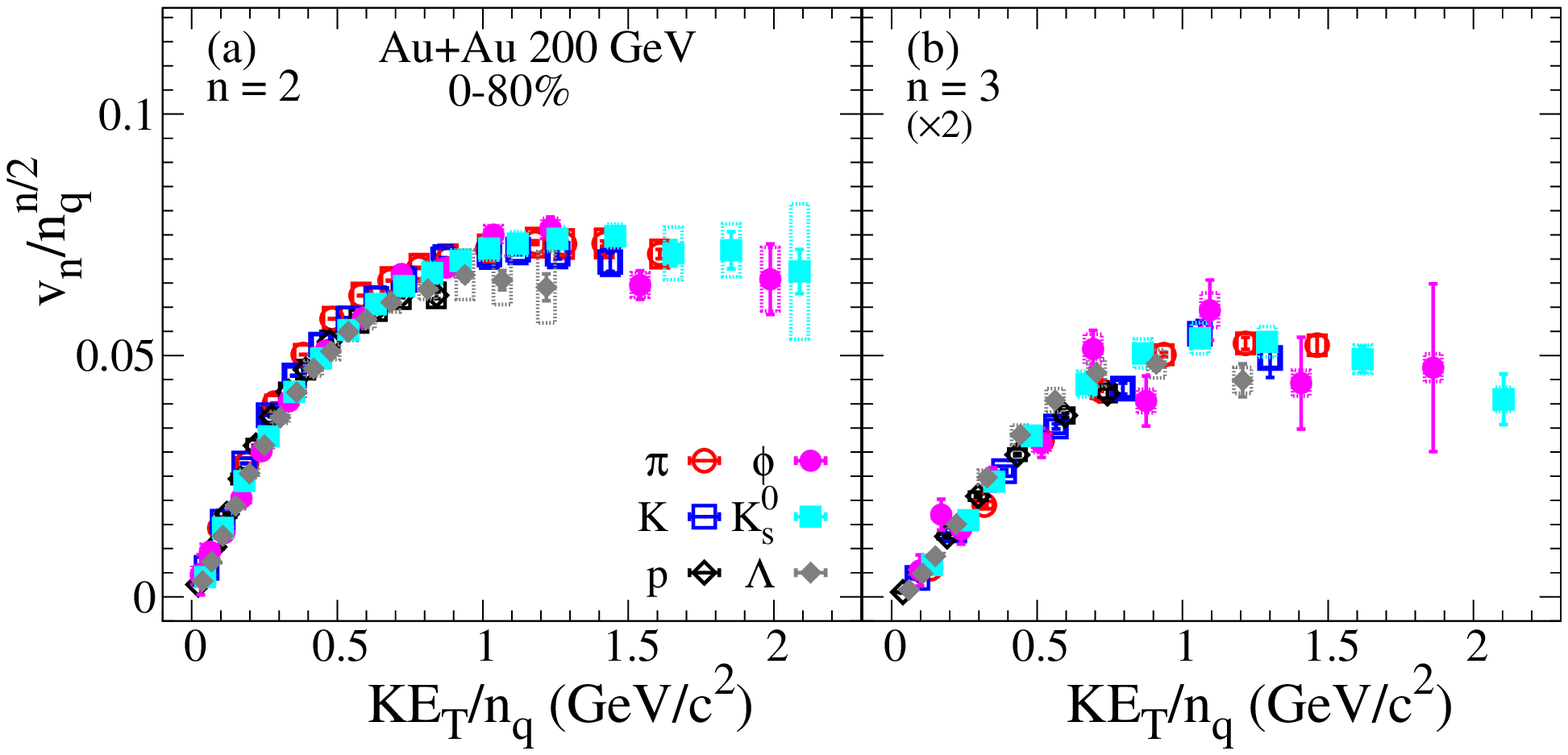}
\vskip -0.36cm
\caption{
The scaled identified particle elliptic and triangular flow versus the scaled transverse kinetic energy for 0--80\% central Au+Au collisions at $\sqrt{s_{\rm NN}}$ = 200 GeV.
\label{fig:fig4}
 }
}
\vskip -0.2cm
\end{figure}

Figure~\ref{fig:fig4} shows the number of constituent quark ($n_q$) scaled $v_2$ (a) and $v_3$ (b) as a function of scaled kinetic energy dependence at midrapidity ($|y|<1.0$) of  $\pi$, $K$, $p$,  $\Lambda$, $\varphi$ and $K_{s}^{0}$ for 0--80\% central collisions. 
The measurements indicate a clear scaling for the $v_{n}/n_{q}^{n/2}$ vs. scaled kinetic energy $KE_{T}/n_{q}$~\cite{Han:2011iy} at the top RHIC energy of $\sqrt{s_{\rm NN}}$ = 200~GeV. The observed scaling properties of $v_{2}$ and $v_{3}$ imply that the measured collective flow develops during the partonic phase. 

\begin{figure*}[th]
\vskip -0.36cm
\centering{
\includegraphics[width=0.99 \linewidth,angle=0]{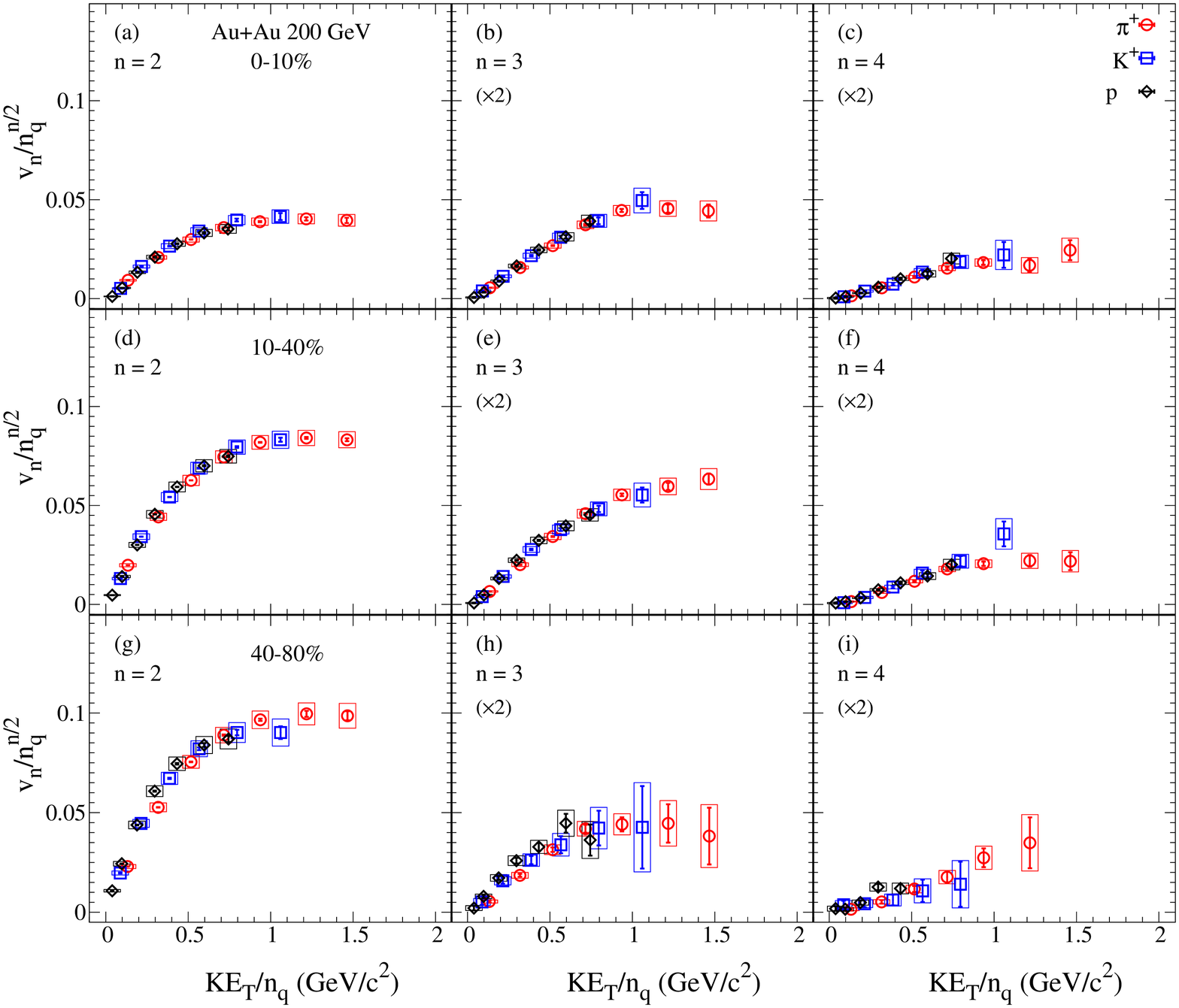}
\vskip -0.36cm
\caption{
The scaled identified particle $v_{2}$, $v_{3}$, and $v_{4}$ versus the scaled transverse kinetic energy for 0-10$\%$,  10-40$\%$ and 40-80$\%$ central Au+Au collisions at $\sqrt{s_{\rm NN}}$ = 200 GeV.
\label{fig:fig5}
 }
}
\vskip -0.2cm
\end{figure*}
\begin{figure*}[th]
\vskip -0.36cm
\centering{
\includegraphics[width=0.80 \linewidth,angle=0]{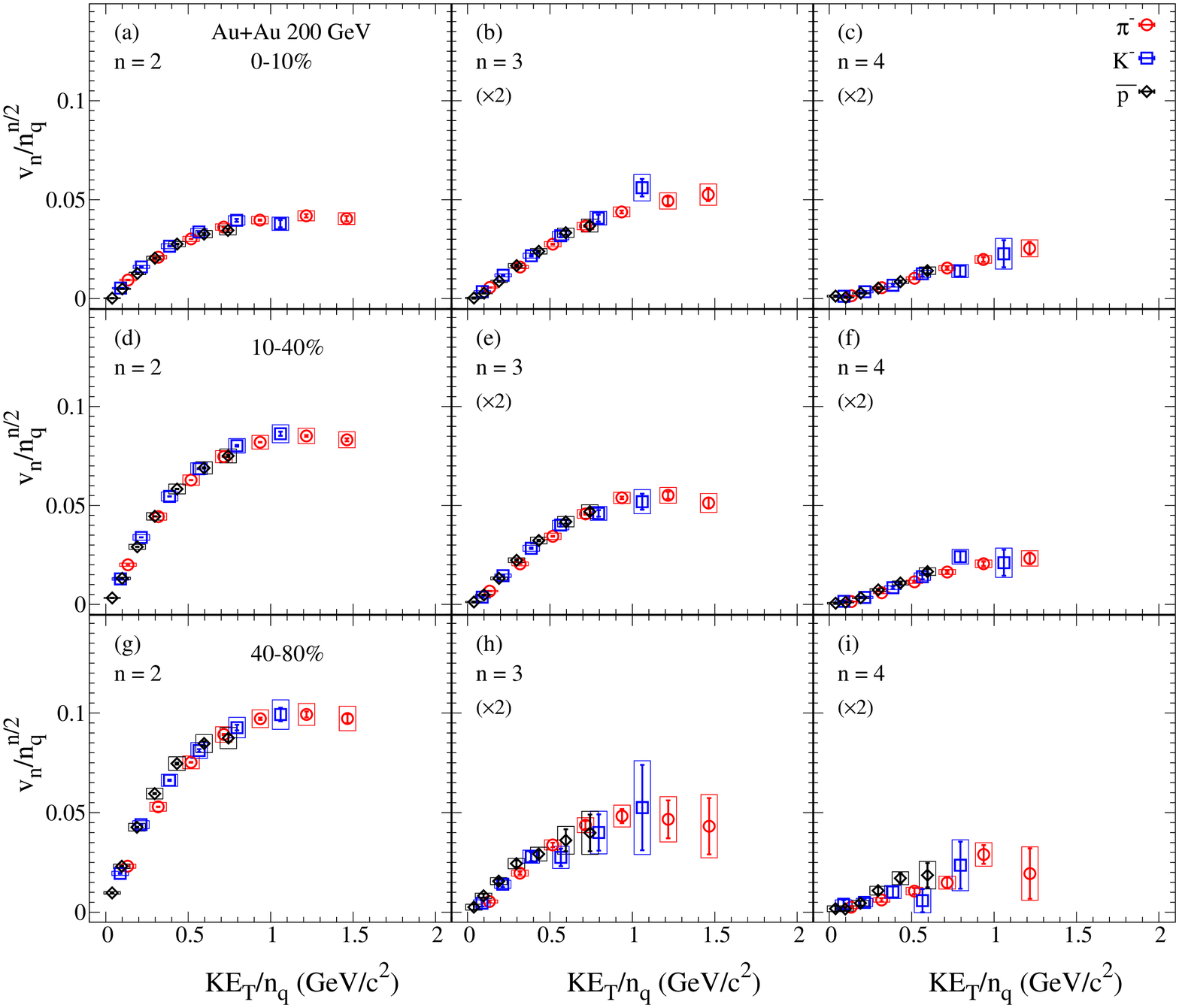}
\vskip -0.36cm
\caption{
The scaled identified anti-particle $v_{2}$, $v_{3}$, and $v_{4}$ versus the scaled transverse kinetic energy for 0--10\%, 10--40\% and 40--80\% central Au+Au collisions at $\sqrt{s_{\rm NN}}$ = 200 GeV.
\label{fig:fig6}
 }
}
\vskip -0.2cm
\end{figure*}
The NCQ scaling properties of these data can be further explored via the centrality dependence of $v_n$ vs. the scaled kinetic energy. Figures~\ref{fig:fig5} and ~\ref{fig:fig6} show  the scaled kinetic energy dependence of $v_2$, $v_3$, and $v_4$  for $\pi^{+}$, $K^{+}$, $p$, and $\pi^{-}$, $K^{-}$, $\bar{p}$  at midrapidity ($|y|<1.0$) of 0--10\%, 10--40\% and 40--80\% central Au+Au collisions. The measurements indicate a scaling of the $v_{n}/n_{q}^{n/2}$ vs. $KE_{T}/n_{q}$~\cite{Han:2011iy} for all the centrality intervals shown. Such measurements could add constraints to theoretical models attempting to reproduce the anisotropic flow.

\subsection{$v_{n}$ as a function of centrality and comparison with models}
\label{Sec:4-3}

\begin{figure*}[t]
\vskip -0.36cm
\centering{
\includegraphics[width=0.60 \linewidth,angle=0]{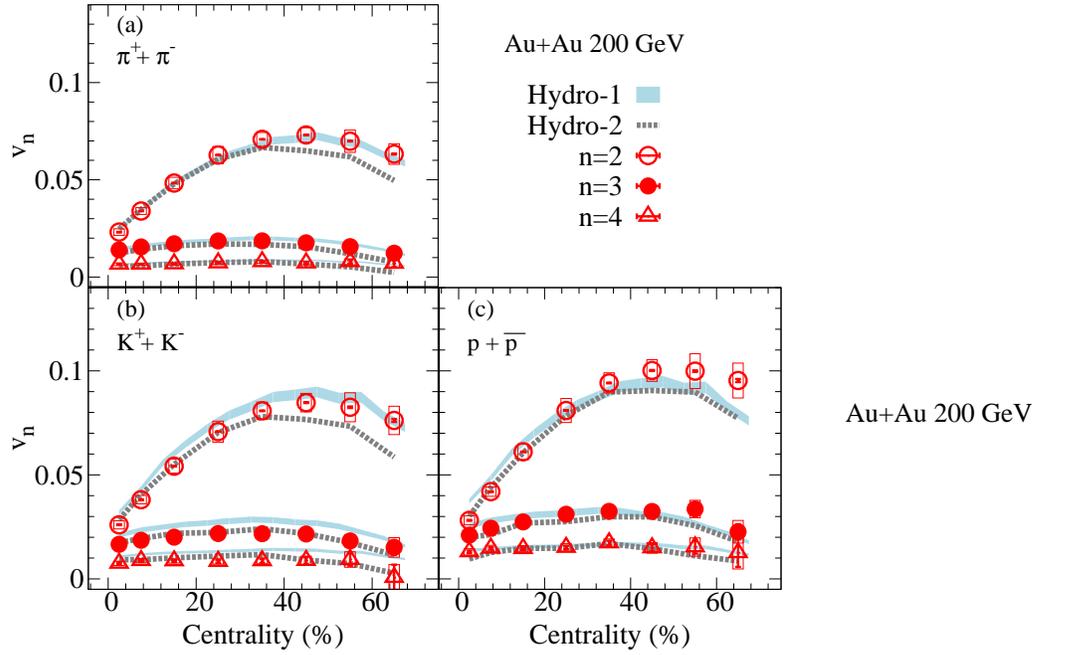}
\vskip -0.36cm
\caption{
The  centrality dependence of the $\pi$, $K$, and $p_T$ integrated $v_{2}$, $v_{3}$ and $v_{4}$ values for $p_{T}$ $<$ 2.0 GeV/c in Au+Au collisions at $\sqrt{s_{\rm NN}}$ = 200 GeV. The solid and dashed lines represent the two hydrodynamic models used here~\cite{Alba:2017hhe,Schenke:2019ruo}.
\label{fig:fig7}
 }
}
\vskip -0.2cm
\end{figure*}

The centrality dependence of the $p_T$-integrated $v_{2}$, $v_{3}$ and $v_{4}$ of $\pi$, $K$, and $p$ for Au+Au collisions at $\sqrt{s_{\rm NN}}$ = 200 GeV are presented in Fig.~\ref{fig:fig7}. 
The flow harmonics show a characteristic dependence on the collision centrality that reflects the interplay between initial-state effects and the final-state effects from central to peripheral collisions~\cite{Liu:2018hjh,STAR:2019zaf}.
In addition, the $v_{n}$ values decrease with increasing harmonic order. Such an observation reflects the increase of viscous effects with increasing harmonic order~\cite{STAR:2019zaf}. {\color{black}The weakening centrality dependence for higher flow harmonic, especially $v_3$, is caused by the dominating geometry fluctuations.}

{\color{black}Thorough comparisons between data and theoretical calculations are carried out for all harmonics $v_{2}$, $v_{3}$, and $v_{4}$.} The shaded bands in Fig.~\ref{fig:fig7} indicate two viscous hydrodynamic model predictions~\cite{Alba:2017hhe,Schenke:2019ruo} which are summarized in Table~\ref{tab:2}. Note that these two models differ in their initial- and final-state assumptions. 
However, both models show qualitative agreement with the present measurements. The predictions from Hydro$-1$ (cf. Table~\ref{tab:2}) give a closer description to the measured $v_{2}$ values.  The Hydro$-1$ model overpredicts the kaon $v_{3}$ and $v_{4}$ values. The Hydro$-2$ model gives a closer description to the $v_{3}$ and $v_{4}$ values. 

\clearpage
\section{Summary}
\label{Sec:5}

In summary, we have presented new differential measurements of $v_{2}$,  $v_{3}$ and $v_{4}$, at midrapidity ($|\eta|<1.0$) for identified hadrons as a function of centrality and transverse momentum in Au+Au collisions at  $\sqrt{s_{\rm NN}}=$ 200~GeV. 
The $p_T$-differential measurements indicate a sizable centrality and mass-order dependence for the measured flow harmonics. The similarities of the shapes of the $v_{n}$ versus $p_T$ curves for light and strange mesons indicate that the heavier s quarks flow as strongly as the lighter u and d quarks.   
We also observed number of consituent quark scaling for $v_2$, $v_3$, and $v_4$ which suggests that the measured collective flow develops during the partonic phase. 
Furthermore, a qualitative agreement between the present flow measurements and the two viscous hydrodynamic calculations was obtained. These comparisons may provide additional constraints on the transport properties of the medium produced in these collisions.

\begin{acknowledgements}
%
We thank the RHIC Operations Group and RCF at BNL, the NERSC Center at LBNL, and the Open Science Grid consortium for providing resources and support.  This work was supported in part by the Office of Nuclear Physics within the U.S. DOE Office of Science, the U.S. National Science Foundation, the Ministry of Education and Science of the Russian Federation, National Natural Science Foundation of China, Chinese Academy of Science, the Ministry of Science and Technology of China and the Chinese Ministry of Education, the Higher Education Sprout Project by Ministry of Education at NCKU, the National Research Foundation of Korea, Czech Science Foundation and Ministry of Education, Youth and Sports of the Czech Republic, Hungarian National Research, Development and Innovation Office, New National Excellency Programme of the Hungarian Ministry of Human Capacities, Department of Atomic Energy and Department of Science and Technology of the Government of India, the National Science Centre of Poland, the Ministry  of Science, Education and Sports of the Republic of Croatia, RosAtom of Russia and German Bundesministerium f\"ur Bildung, Wissenschaft, Forschung and Technologie (BMBF), Helmholtz Association, Ministry of Education, Culture, Sports, Science, and Technology (MEXT) and Japan Society for the Promotion of Science (JSPS).

\end{acknowledgements}

\appendix

\section{Scaled $v_{n}$ as a function of scaled $p_T$}\label{A1}
\begin{figure}[th]
\vskip -0.36cm
\centering{
\includegraphics[width=0.70 \linewidth,angle=0]{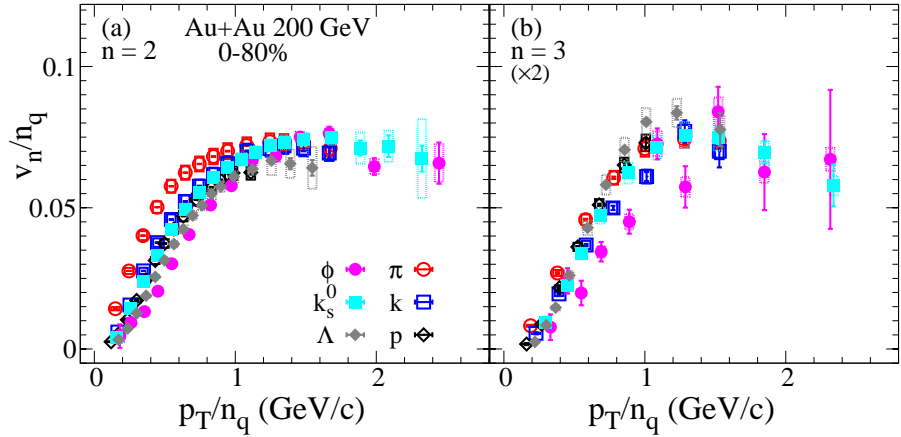}
\vskip -0.36cm
\caption{
The scaled identified particle elliptic and triangular flow versus the scaled $p_T$  for 0--80\% central Au+Au collisions at $\sqrt{s_{\rm NN}}$ = 200 GeV.
\label{fig:A11}
 }
}
\vskip -0.2cm
\end{figure}

\begin{figure*}[th]
\vskip -0.36cm
\centering{
\includegraphics[width=0.99 \linewidth,angle=0]{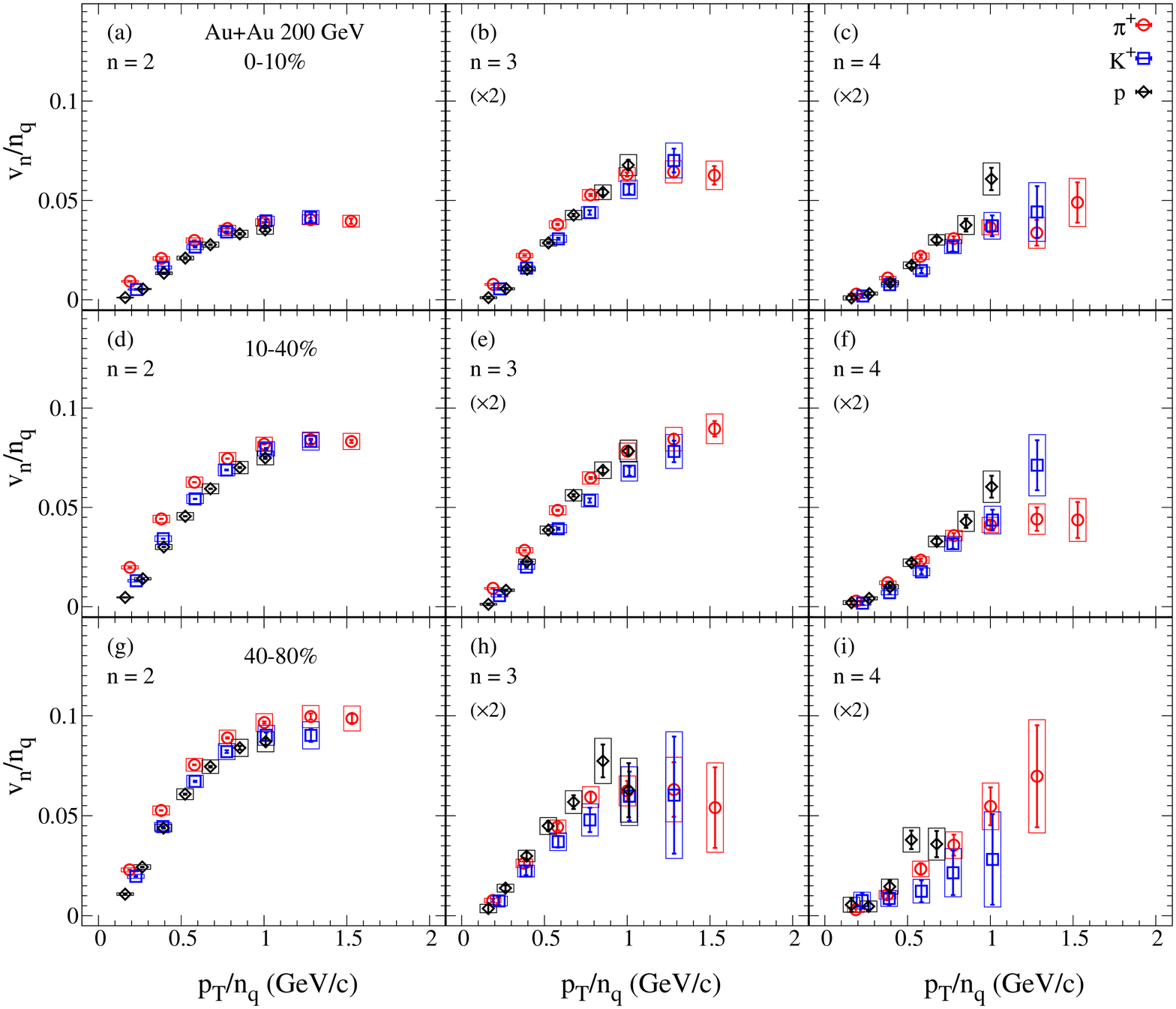}
\vskip -0.36cm
\caption{
The scaled identified particle $v_{2}$, $v_{3}$, and $v_{4}$ versus the scaled $p_T$ for 0-10$\%$,  10-40$\%$ and 40-80$\%$ central Au+Au collisions at $\sqrt{s_{\rm NN}}$ = 200 GeV.
\label{fig:A12}
 }
}
\vskip -0.2cm
\end{figure*}

\begin{figure*}[th]
\vskip -0.36cm
\centering{
\includegraphics[width=0.99 \linewidth,angle=0]{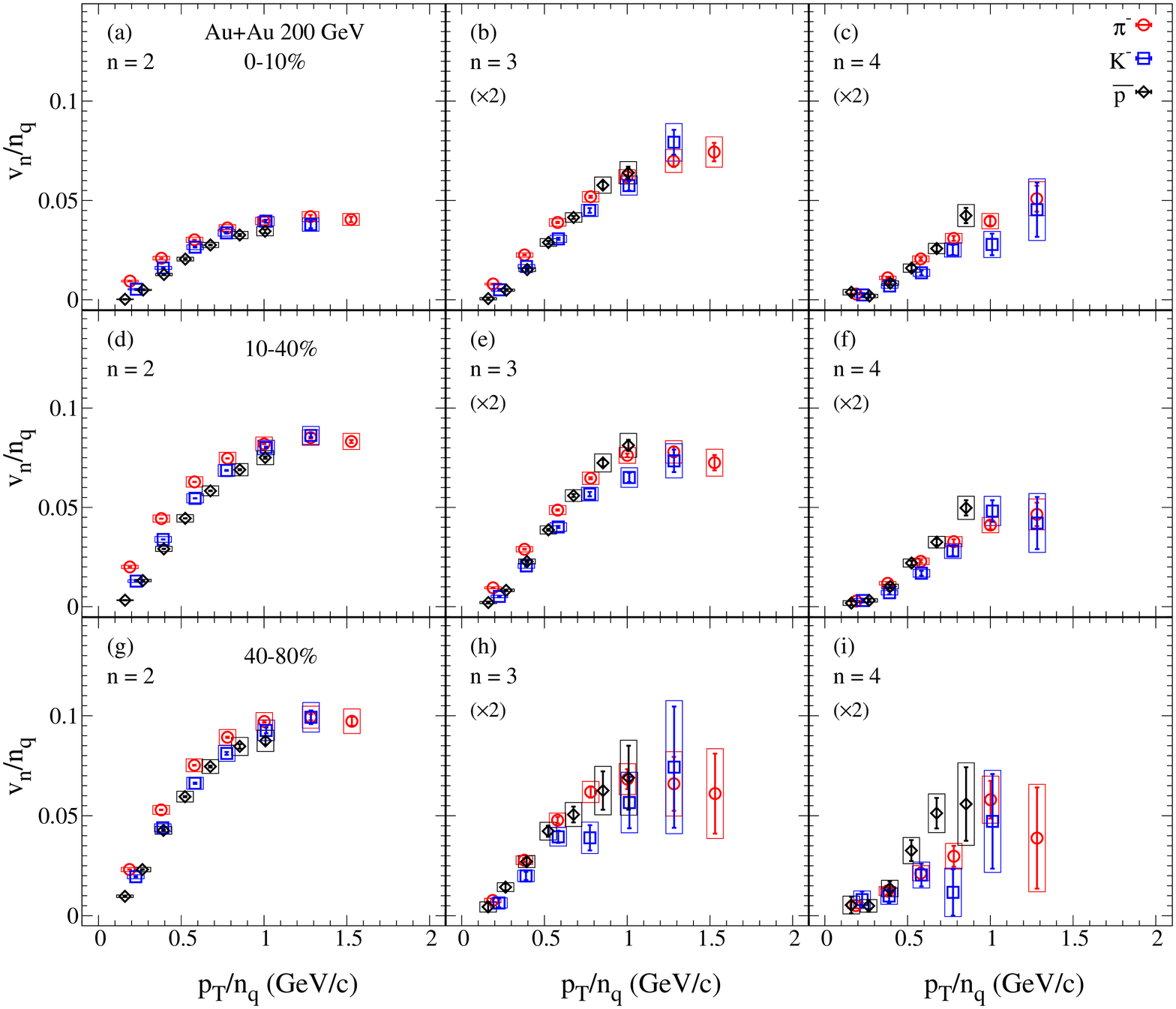}
\vskip -0.36cm
\caption{
The scaled identified anti-particle $v_{2}$, $v_{3}$, and $v_{4}$ versus the scaled $p_T$ for 0--10\%, 10--40\% and 40--80\% central Au+Au collisions at $\sqrt{s_{\rm NN}}$ = 200 GeV.
\label{fig:A13}
 }
}
\vskip -0.2cm
\end{figure*}

The number of constituent quark scaling~\cite{STAR:2017ykf,PHENIX:2014uik,STAR:2021twy} can be employed to show the collective flow was generated  at the partonic level. In the number of constituent quark scaling process, at a given $p_T$ hadrons are created from $n_q$ quarks with transverse momentum $p_T$/$n_q$. 
Figures~\ref{fig:A11},~\ref{fig:A12} and~\ref{fig:A13} presents $v_n$/$n_q$ of different particle species as a function of $p_T$/$n_q$.
The number of constituent quark scaled $v_n$ as a function of $p_T$/$n_q$ seems to show a global tendency for all particles species, although there are small differences for each $v_n$.


\bibliography{ref} 
\end{document}